\PassOptionsToPackage{unicode}{hyperref}
\PassOptionsToPackage{hyphens}{url}
\documentclass[]{article}
\usepackage{amsmath,amssymb}
\usepackage{lmodern}
\usepackage{iftex}
\ifPDFTeX
  \usepackage[T1]{fontenc}
  \usepackage[utf8]{inputenc}
  \usepackage{textcomp} % provide euro and other symbols
\else % if luatex or xetex
  \usepackage{unicode-math}
  \defaultfontfeatures{Scale=MatchLowercase}
  \defaultfontfeatures[\rmfamily]{Ligatures=TeX,Scale=1}
\fi
\IfFileExists{upquote.sty}{\usepackage{upquote}}{}
\IfFileExists{microtype.sty}{% use microtype if available
  \usepackage[]{microtype}
  \UseMicrotypeSet[protrusion]{basicmath} % disable protrusion for tt fonts
}{}
\makeatletter
\@ifundefined{KOMAClassName}{% if non-KOMA class
  \IfFileExists{parskip.sty}{%
    \usepackage{parskip}
  }{% else
    \setlength{\parindent}{0pt}
    \setlength{\parskip}{6pt plus 2pt minus 1pt}}
}{% if KOMA class
  \KOMAoptions{parskip=half}}
\makeatother
\usepackage{xcolor}
\IfFileExists{xurl.sty}{\usepackage{xurl}}{} % add URL line breaks if available
\IfFileExists{bookmark.sty}{\usepackage{bookmark}}{\usepackage{hyperref}}
\hypersetup{
  hidelinks,
  pdfcreator={LaTeX via pandoc}}
\ifLuaTeX
  \usepackage{selnolig}  % disable illegal ligatures
\fi

\author{}
\date{}

\begin{document}

\title{{\LARGE History and Problems of the Standard Model in Cosmology}}

\author{{\Large Mart\'\i n L\'opez-Corredoira}\footnote{Instituto de Astrof\'\i sica de Canarias,
E-38205 La Laguna, Tenerife, Spain;\\
PIFI-Visiting Scientist 2023 of Chinese Academy of Sciences at Purple Mountain Observatory, 210023 Nanjing (China) and National Astronomical Observatories, 100012 Beijing (China);\\
Departamento de Astrof\'\i sica, Universidad de La Laguna,
E-38206 La Laguna, Tenerife, Spain;\\
martin@lopez-corredoira.com}}

\maketitle

Since the beginning of the 20th century, a continuous evolution and
perfection of what we today call the standard cosmological model has
been produced, although some authors like to distinguish separate
periods within this evolution. A possible historical division of the
development of cosmology into six periods was proposed by Luminet
(2008): (1) the initial period (1917--1927); (2) the period of
development (1927--1945); (3) the period of consolidation (1945--1965);
(4) the period of acceptance (1965--1980); (5) the period of enlargement
(1980--1998), and (6) the period of high-precision experimental
cosmology (1998--).

\section{The Initial Period (1917-1927)}

At the beginning of the 20th century, two great achievements in physics
and astronomy initiated the journey toward the standard cosmological
model as we know it today. The first was the observational evidence for
the existence of many galaxies separated by very large distances---much
larger than the usual distances managed by astronomers previously---the
Milky Way thus being only one galaxy among many. It was deﬁnitively
established after a period of discussion that ﬁnished with the Great
Debate in 1920 between the American astronomers Heber D Curtis
(1872--1942), who defended the hypothesis that some nebulae (now called
galaxies) were not part of the Milky Way but were located at very large
distances from it, and Harlow Shapley (1885--1972), who claimed that
these nebulae were part of the Milky Way. This achievement gave rise to
the subsequent development of extragalactic astronomy, and, implicitly,
a new cosmological vision was emerging out of this scenario: a vision of
a Universe of vast spaces, impossible to imagine, where galaxies are the
fundamental components in a larger-scale structure.

The other great achievement came from physics in the form of Albert
Einstein's (1879--1955) general relativity. Certainly, his earlier
theory of special relativity was also very important, but for astronomy,
particularly from the perspective of cosmology, general relativity was
the long-awaited breakthrough. Newton's magniﬁcent achievements had
blocked the free expansion of cosmological ideas because of the problems
in solving the stability of systems without an eventual collapse and
having recourse to godly intervention.

The models that would constitute the basis of our present standard
cosmology came a little later. The basic idea assumed is that the
current Universe is homogeneous on a large scale and that the distances
among all the different objects are currently growing owing to the
expansion of the Universe, a recession of objects with respect to one
another on a large scale. On small scales, different objects could
cluster together because their gravitational attraction overcomes the
expansion. The Russian physicist Alexander Friedmann (1888--1925)
developed the basic aspects of the application of general relativity to
a cosmological model (Friedmann 1922, 1924).\\

\section{The Period of Development (1927--1945)}

In 1924, the German astronomer Carl Wirtz (1876--1939) noted a
correlation between the faintness of a galaxy and its redshift. Edwin P
Hubble (1889--1953) and Milton Humason (1891--1972) measured the
distance of a number of galaxies during the same year and would later
ﬁnd the famous Hubble--Lema\^itre law of the linear relationship between
radial velocities and distances. The redshifts were interpreted as proof
of the expansion of the Universe (Hubble, 1929). Prior to Hubble's
publication in 1927, the Belgian Catholic priest, physicist, and
astronomer Georges Lema\^itre (1894--1966) developed a theoretical model
of an expanding Universe in an extension of the work of Friedmann. The
work by Lema\^itre (1927) was published in French in a small Belgian
journal, and also tells us about the recession of galaxies and the
recession rate in the linear velocity--distance relationship, including
an analysis of observational data, as rediscovered later by Hubble in
1929.

Another line of development of the cosmological model was suggested by
the Japanese physicist Seitaro Suzuki, who suggested that the observed
helium--hydrogen ratio might be explained ``if the cosmos had, at the
creation, the temperature higher than 109 degrees'' (Suzuki, 1928).
Lema\^itre, in 1931, with the expansion and the arrow of time from the
second law of thermodynamics in mind, developed his concept of the
`primeval atom' (Lema\^itre, 1931), the ﬁrst version of what later would
be called the ``Big Bang''. According to him, the initial state of
matter in the Universe might be thought of as a sea of neutrons.
Lema\^itre thought that cosmic rays were relics of primordial decays of
atoms, which was later demonstrated to be wrong. Moreover, his ideas on
stellar evolution were also demonstrated to be wrong during the 1930s.
So, by the end of the decade, the primeval-atom hypothesis had been
generally rejected by the scientiﬁc community.

\section{The Period of Consolidation (1945--1965)}

After World War II, George Gamow (1904--1968), a Russian physicist who
emigrated to the US in 1934, compared the detonation of an atomic bomb
with the origin of the Universe and popularized the Big Bang theory
(Gamow, 1947). In fact, the name ``Big Bang'' was not given by Gamow,
but by one of the opponents of his theory, Fred Hoyle (1915--2001), who
dubbed Gamow's primeval atom theory as the ``Big Bang'', in order to
ridicule it. Gamow and one of his students, Ralph Alpher (1921--2007),
published a paper in 1948. Gamow, who had a certain sense of humor,
decided to put the reputed physicist Hans Bethe (1906--2005) as the
second author, even though he had not participated in the development of
the paper, so the result was a paper by Alpher, Bethe, and Gamow (Alpher
et al., 1948), to rhyme with ``alpha, beta and gamma''. Later, Robert
Herman (1914--1997) joined the research team, but---according to
Gamow---he stubbornly refused to change his name to ``Delter.''

Alpher and Herman (1949) and Gamow (1953) also predicted an early stage
of the Universe that would produce relic radiation that could be
observed at present as a background in microwave wavelengths,
corresponding to the epoch of decoupling of matter and radiation. The
ﬁrst published recognition of relic radiation as a detectable microwave
phenomenon was in 1964 by the Russian cosmologists Andrei Doroshkevich
(1937--) and Igor Dmitriyevich Novikov (1935--) (Doroshkevich \&
Novikov, 1964). Then came the ofﬁcial discovery of the cosmic microwave
background radiation by Arno Allan Penzias (1933--) and Robert Woodrow
Wilson (1936-) (Penzias \& Wilson, 1965), although this same radiation
had been previously directly or indirectly observed by other
researchers.

\section{The Period of Acceptance (1965--1980)}

More evidence supporting the standard model of the expanding Universe
came from Malcolm Longair (1941--) and Martin Ryle (1918--1984), who
argued that the data indicate that the Universe must be evolving
(Longair, 1966; Ryle, 1968). The galaxies at high redshift---that is, at
great distance---showed distributions and properties different from
those at low redshift. Since at larger distances we are observing the
past Universe, given the limited speed of light, this implies that the
distant galaxies belong to an epoch of the Universe that was much
earlier than the present one.

The conﬁrmation of the predicted microwave radiation and evolution of
the Universe gave conﬁdence to those cosmologists who supported the
standard model. Many hitherto skeptical physicists and astronomers
became convinced they now had a solid theory. By the mid-seventies,
cosmologists' conﬁdence was such that they felt able to describe in
intimate detail events of the ﬁrst minutes of the Universe (Weinberg,
1977).\\

\section{The Period of Enlargement (1980--1998)}

Nonetheless, there were problems that remained to be solved, such as why
the Universe appeared to be the same in all directions (isotropic), why
the cosmic microwave background radiation was evenly distributed, and
why its anisotropies were so small. Why was the Universe ﬂat and the
geometry nearly Euclidean? How did the large-scale structure of the
cosmos originate? Clearly, work on the fundamental pillars of the
cosmological ediﬁce remained to be done. In the 1970s and 1980s,
proposals were brought forth to solve these pending problems, with
inﬂation as the leading idea in the solution of cosmological problems at
the beginning of the Universe, and the idea of non-baryonic dark matter
as a new paradigm that allows the theory to ﬁt the numbers of some
observations. Grand Uniﬁed Theories of particle physics would also
support the existence of CP violation (asymmetry of matter and
antimatter) or non-baryonic dark matter. Also, the joining of cosmology
and particle physics and scenarios containing baby universes, wormholes,
superstrings, and other exotic ideas were born. This excess of
theoretical speculation, not based on observations, has led some authors
to call this epoch the era of post-modern cosmology (Bonometto, 2001).
This union between cosmology and particle physics is due in part to the
halting of particle physics experiments because of their escalating
cost, a situation that led many particle physicists to move over into
cosmology, wishfully contemplating the Universe as the great accelerator
in the sky (Disney, 2000; White, 2007). Alas, particle physicists lack
the necessary astronomical background---complained Mike Disney---to
appreciate how soft an observational, as opposed to experimental
science, necesarily has to be.

In the 1990s, a third patch was applied to the theory in an effort to
solve new inconsistencies with the data in the form of dark energy,
which supposedly produced acceleration in the cosmic expansion. The
problems to be solved were basically the new Hubble--Lema\^itre diagrams
with type Ia supernovae as putative standard candles, the numbers
obtained from cosmic microwave background radiation anisotropies, and
especially estimates of the age of the Universe, which were inconsistent
with the calculated ages of the oldest stars.

The renovated standard model, including these ad hoc elements, would
come to be called the $\Lambda $CDM cosmological model, where $\Lambda $ stands for dark
energy, and CDM stands for cold dark matter, the favored subgroup of
models of non-baryonic dark matter. Some cosmologists referred to it as
`concordance cosmology' to emphasize that this model is in agreement
with all the known observations. Other authors, critical of the standard
model, prefer to call it `consensus cosmology,' wishing to emphasize
that this new cosmology is, above all, a sociological question of
agreement among powerful scientiﬁc teams in order to establish the
orthodoxy of a fundamental dogma. This agreement would be mainly between
two powerful cosmological groups, the teams dedicated to the analysis of
supernovae and the cosmic microwave background, who found a rough
coincidence in the necessary amount of dark energy, although with large
error bars, that reinforced their belief that they had discovered an
absolute truth, thus compelling the rest of the community to accept this
truth as a solid standard, while at the same time discarding the results
of other less powerful cosmological groups that presented different
values of the parameters. Talking about consensus cosmology, Rudolph
(`Rudy') Schild (1940--) once queried, ``Which consensus? Do you know
who consented? A bunch of guys at Princeton who drink too much tea
together''\,' (Unzicker \& Jones, 2013, ch. 3).

\section{The Period of High-Precision Experimental Cosmology (1998--)}

Rather than major discoveries or proposals, this epoch is characterized
by a lack of discussion on the fundamental ideas in cosmology, when it
becomes a tenet of belief that all the major problems have been solved.
This state of complacency has resulted in excess conﬁdence in the
robustness and superiority of the standard model, with little
consideration for alternative models. Certainly, some minor topics are
being debated, such as the equation of the state of dark energy, and the
types of inﬂation or the coldness or hotness of dark matter, but these
are subtleties (Byzantine arguments) within the major fundamental
scheme. This is the epoch in which the main enterprise of cosmology
consists of spending big money on megaprojects that will achieve
accurate measurements of the values of the cosmological parameters and
solve any small problems that remain to be explained.

This is also the epoch of the highest social recognition of cosmology:
Not only do schools, museums, and popular science journals talk about
the Big Bang as well established, to be compared to Darwin's evolution
and natural selection theory, but cosmology now occupies a privileged
ranking among the most prestigious natural sciences. For instance,
cosmology and its four dark knights (CP violation, inflation,
non-baryonic dark matter, and dark energy) have been awarded Nobel
Prizes in Physics in 2011 and 2019, respectively, for the putative
discovery of the dark energy that produces the acceleration of the
expansion, and the inclusion of the dark components in our understanding
of the Universe. One may wonder whether unconﬁrmed quasi-metaphysical
speculations should properly form part of the body of the recognized
knowledge of physics, leaving behind the conservative tradition of Nobel
committees not awarding prizes for speculative proposals. Einstein did
not receive either of his Nobel Prizes for his discovery of special and
general relativity; neither did Curtis for his deﬁnitive recognition of
the true nature of galaxies in the Great Debate of 1920. Neither
Lema\^itre nor Hubble received the Nobel Prize for their discovery of the
expansion of the Universe, but we now have committees that give maximum
awards for highly speculative proposals, such as the acceleration of the
expansion of the Universe, the reality of which has yet to be conﬁrmed.
We certainly do live in a very special time for cosmology.

However, this brand of epistemological optimism has declined with time,
and the expression ``crisis in cosmology'' is stubbornly reverberating
in the media. The initial expectation of removing the pending minor
problems arising from the increased accuracy of measurements has
backﬁred: the higher the precision with which the standard cosmological
model tries to ﬁt the data, the greater the number of tensions that
arise, the problems proliferating rather than diminishing. Moreover,
there are alternative explanations for most of the observations.

At the Anomalies in Modern Astronomy Research online symposium organized
by the Society of Scientific Exploration (October 22nd, 2022), Prof.
Pavel Kroupa presented anomalies from galactic to Gpc scales
(large-scale structures), including some examples of 5$\sigma $ tensions and
some mention of Modified Newtonian dynamics (MOND) as an alternative to
standard gravity and dark matter. We can complement the range of
anomalies in cosmology with further cases of Cosmic Microwave Background
Radiation, nucleosynthesis, tests of expansion, CP violation, inflation,
and other topics. There is no space in the present text to discuss in
detail these topics; the reader interested in these anomalies and
tensions can read the recent literature on the collections of problems
of the standard model: (Perivolaropoulos \& Skara, 2022; Abdalla et al.,
2022; Melia, 2022; L\'opez-Corredoira, 2017, 2022).

CP violation has problems; There is no experimental evidence for a ﬁnite
lifetime of a proton below 10$^{34}$ years (Tanaka et al., 2020). Inflation
has problems; Some authors have argued that the inﬂation necessary to
explain a ﬂat Universe is highly improbable (Iljas et al., 2017).
Hubble--Lema\^itre diagrams with type Ia supernovae can be explained
without dark energy (L\'opez-Corredoira \& Calvo-Torel, 2022); also, dark
energy can be avoided in other observations.

The standard interpretation of the redshifts of galaxies is that they
are due to the expansion of the Universe plus peculiar motions, but
there are other explanations, such as the ``tired light'' hypothesis,
which assumes that the photon loses energy owing to some unknown
photon--matter process or photon--photon interaction when it travels
some distance. Different observational tests give different results,
although none of them so far provides strong proof in favor of a static
Universe (L\'opez-Corredoira, 2017; L\'opez-Corredoira, 2022, ch. 4). The
discussion on anomalous redshifts is also inconclusive.

Doubt is cast upon that precision cosmology derived from Cosmic
Microwave Background Radiation analysis, owing to the difﬁculties in
making maps totally free from foreground contamination. Moreover, many
alternative explanations of its origin are found in the technical
literature, and certain observed anomalies, such as the lack of low
multipole signal, alignment of quadrupole and octupole, and others, are
at odds with the standard model (Schwarz et al., 2016), which opens the
door to possible fundamental errors in the standard cosmological
description of this radiation.

In the standard model, it is claimed that helium-4, lithium-7, and other
light elements were created in the primordial Universe, and the
existence of these elements was used as proof for the necessity of a hot
Universe in its ﬁrst minutes of life. However, only helium-4 has had
successful direct conﬁrmation of the predictions, although at the price
of requiring a baryon density raises other problems. The observed
abundance of lithium-7 is 3 to 4 times lower than predicted (Coc et al.,
2012). The other light elements are affected by uncertainties in the
theoretical model or by later creation or destruction associated with
stellar nucleosynthesis, cosmic rays, or other astrophysical processes,
so they cannot be used to corroborate cosmological predictions.
Moreover, there are alternatives to primordial nucleosynthesis to
explain the observed abundances, even for helium-4 (Adouze et al., 1985;
Burbidge \& Hoyle, 1998).

Cosmology is not a science like others since it contains more
speculative elements than is usual in other branches of physics, with
the possible exception of particle physics. The goal of cosmology is
also more ambitious than routine theories in physics: cosmology aims to
understand everything in our Universe without limit. However,
cosmological hypotheses should be very cautiously proposed and even more
cautiously received. This skepticism is well-founded. There are
scientiﬁc, philosophical, and sociological arguments to support this
claim (L\'opez-Corredoira, 2022).

\emph{Some of the material for this article was excerpted from the
book {\rm Fundamental Ideas in Cosmology. Scientific, Philosophical and
Sociological Critical Perspectives} (L\'opez-Corredoira, 2022). The author
 is supported by the Chinese Academy of Sciences President’s 
 International Fellowship Initiative grant number 2023VMB0001.}

\end{document}